\documentclass[useAMS,usenatbib]{mn2e}
\usepackage{graphics}
\usepackage{rotating}
\usepackage{multirow}

\title[Microlensing by a wide-separation planet]
{Microlensing by a wide-separation planet: detectability and
boundness}
\author[Ryu et al.]
{Yoon-Hyun Ryu$^{1}$\thanks{E-mail: yoonhyunryu@gmail.com(YHR);
mgp@knu.ac.kr (MGP); hyc@knu.ac.kr (HYC); leekw@cu.ac.kr (KWL)},
Myeong-Gu Park$^1$\footnotemark[1], Heon-Young
Chang$^1$\footnotemark[1],
and Ki-Won Lee$^2$\footnotemark[1]\\
$^{1}$Department of Astronomy and Atmospheric Sciences, Kyungpook
National University, Daegu 702-701, Rep. of Korea\\
$^{2}$Institute of Liberal Education, Catholic University of Daegu,
Gyeongbuk 712-702, Rep. of Korea}
\begin{document}

\maketitle

\label{firstpage}

\begin{abstract}

The microlensing technique has proven sensitive to Earth-like and/or
wide-separation extrasolar planets. The unbiased spatial
distribution of extrasolar planets with respect to host stars is
crucial in studying the planet formation processes. If one can
characterize the planetary microlensing light curves whether they
are produced by a wide-separation planet or a free-floating planet,
it will greatly help to establish the spatial distribution of
extrasolar planets without contamination by free-floating planets.
Previous studies have shown that the effect of the host star on the
microlensing by the accompanying wide-separation planet can be
significant enough to be detected by the high-frequency microlensing
experiments for typical microlensing parameters. Here, we further
explore the detection condition of a wide-separation planet through
the perturbation induced by the planetary caustic for various
microlensing parameters, especially for the size of the source
stars. By constructing the fractional deviation maps at various
positions in the space of microlensing parameters, we find that the
pattern of the fractional deviation depends on the ratio of the
source radius to the caustic size, and the ratio satisfying the
observational threshold varies with the star-planet separation. We
have also obtained the upper limits of the source size that allows
the detection of the signature of the host star as a function of the
separation for given observational threshold. It is shown that this
relation further leads one to a simple analytic condition for the
star-planet separation to detect the boundness of wide-separation
planets as a function of the mass ratio and the source radius. For
example, when 5$\%$ of the detection threshold is assumed, for a
source star with the radius of $\sim1$ $R_{\odot}$, an Earth-mass
planet and a Jupiter-mass planet can be recognized of its boundness
when it is within the separation range of $\sim10$ AU and $\sim30$
AU, respectively. We also compare the separation ranges of detection
by the planetary caustic with those by the central caustic. It is
found that when the microlensing light curve caused by the planetary
caustic happens to be analyzed, one may afford to support the
boundness of the wide-separation planet farther than when that
caused by the central caustic is analyzed. Finally, we conclude by
briefly discussing the implication of our findings on the
next-generation microlensing experiments.
\end{abstract}

\section{INTRODUCTION}

Up to the present, 19 extrasolar planets have been discovered
towards the Galactic bulge by the microlensing experiments operating
in the survey and follow-up mode
\citep{bon04,uda05,bea06,gou06,gau08,ben08,don09,sum10,jan10,miy11,
bat11,Mur11,yee12,ben12,bac12,han13,kai13}. A planetary microlensing
event shows a perturbation, lasting for a short-duration of $t_{\rm
E,p} \sim \sqrt{q} t_{\rm E}$, typically about a day and a few hours
for the mass ratio $q$ corresponding to a Jupiter-mass planet and an
Earth-mass planet, respectively, to the standard light curve induced
by a lens star, whose typical timescale is the Einstein timescale
for the lens system $t_{\rm E}$ \citep{mao91,gou92}. Continuous
monitoring with a high cadence is, therefore, prerequisite to avoid
missing planetary microlensing events.

To monitor a wide field of view continuously with high cadence,
existing microlensing experiments, such as, the OGLE collaboration
and MOA collaboration, have recently changed the camera to expand
the field of view or planned to upgrade the telescope.
Next-generation surveys, with a wider field of view and higher
cadence such as Korea Microlensing Telescope Network (KMTNet) and
\textit{Wide-Field Infrared Survey Telescope} (\textit{WFIRST}) are
being prepared to search the extrasolar planets by the microlensing
technique in ground and space, respectively. The main aim of KMTNet
project is to find sub-Earth-mass planets using three 1.6 m wide
field (2 $\times$ 2 degrees field of view) optical telescopes
located in Chile, South Africa, and Australia \citep{kim10}. The key
goal of \textit{WFIRST} mission is to discover planets down to 0.1
Earth-mass in the separation range wider than 0.5 AU, consequently
to dispense a complete statistical census of the planet in our
Galaxy \citep{ben11}.

Such surveys with a frequent sampling can be sufficiently sensitive
to wide-separation and free-floating planets, and thus enable us to
detect wide-separation planets as well as free-floating planets
which are otherwise very difficult to find by other planet search
techniques, such as, radial velocity technique, transit technique,
direct imaging, or pulsar timing analysis
\citep{ben02,han03,sum11,ben12}. In fact, \citet{sum11} recently
reported the discovery of unbound or possibly distantly orbiting
populations with a planetary-mass, based on microlensing survey
observations on MOA-II phase between 2006 and 2007. According to
their statistical estimates the number of such populations is
approximately twice that of main-sequence stars in the Galaxy.

If one can tell via microlensing observations whether these planets
are still bound to their host stars or freely floating without host
stars, essential information can be provided on the spatial
distribution of extrasolar planets with respect to host stars
without contamination by unbound planets so that the core accretion
theory can be tightly constrained \citep{lau04,ida05,ken06,ken08}. A
free-floating planet itself is also important in constraining the
planet formation processes since it is related to the processes,
such as star-planet scattering \citep{hol99,mus05,mal11},
planet-planet scattering \citep{ras96,wei96,lin97,for08,ver12}, and
star death \citep{ver11}.

Several suggestions have been made to distinguish wide-separation
planets from free-floating planets
\citep{ste99a,ste99b,han03,han05,han09b,ste12a,ste12b}. Basically, a
planetary microlensing event can be inferred that it is caused by a
wide-separation planet if the light curve shows any signature of the
host star, such as, a long-term bump, blended light, or a signal of
the planetary caustic in the light curve. The former two signatures
depend on the source trajectory and lens flux, respectively, and
need additional long-term observations for confirmation, while the
last signature will become important in the next-generation surveys
with high survey monitoring frequency. \citet{han03} investigated
the microlensing properties of wide-orbit planets and found that the
signature of the central star can be detected for a large fraction
of Jupiter mass planetary system for a typical source size and
star-planet distance. \citet{han05} comprehensively discussed these
three methods for the detection of a host star, and found that
one-third of all events should show signatures of its host star
regardless of the planet separation through these methods.

In this paper we investigate the condition in terms of lensing
parameters, particularly, the size of source stars, under which
signatures of the host star in the planetary microlensing light
curve can be detected. We construct magnification maps for various
source radii by the inverse ray-shooting method, taking the limb
darkening into consideration, since the planetary caustic located
close to the wide-separation planet is so small that the finite
source effect becomes critical for a given photometric accuracy. As
a result, we obtain a general empirical formula for the upper limit
on the ratio of source size to the planetary caustic size that
allows to detect the signature of the host star as a function of the
separation, which asymptotically approaches to a constant as the
separation goes to infinity, as approximated by \citet{cha79,cha84}.
We also compare the separation ranges to detect the boundness of the
planet to the host star through the channel of the central and
planetary caustics.

This paper is organized as follows. We briefly describe properties
of a planetary caustic for a wide-separation planetary system in
\S2. We discuss how we construct the fractional deviation maps in
\S3. We present conditions to detect signatures of a host star in
the planetary microlensing light curve in \S4. Finally, we summarize
and conclude our results in \S5.

\section{PROPERTIES OF PLANETARY CAUSTIC}

A planetary perturbation occurs when a source star crosses or passes
by the caustic resulted from a star-planet system. Provided that an
extrasolar planetary system has a single planet, the planetary
system produces one, two or three disconnected caustics depending on
the separation between a planet and its host star. In a
wide-separation planetary system, in which the projected separation
of the planet and the host star is larger than the upper limit of
the lensing zone, two disconnected caustics are generated: central
and planetary caustics. Here, the position of the planetary caustic
is so close to the planet that sometimes the planetary caustic is
located within the Einstein ring due to the planet.

The planetary caustic structure can be expressed by a analytic form
\citep[e.g.,][]{wit95, boz00}. The formulae of x$-$ and
y$-$components of the caustics in the polar coordinate system,
$x(\theta)$ and $y(\theta)$, are given by
\begin{eqnarray}
x(\theta)&=&(s-\frac{m_1}{s})+(\varepsilon_1-\frac{m_2}{\varepsilon_1})\cos\theta
+\frac{m_1\varepsilon_1\cos\theta}{s^2}\\
&+&(1+\frac{m_2}{\varepsilon_1^2})\varepsilon_2\cos\theta
+\frac{m_1}{s^3}(\varepsilon_2s\cos\theta-\varepsilon_1^2\cos2\theta),\nonumber
\end{eqnarray}
\begin{eqnarray}
y(\theta)&=&(\varepsilon_1-\frac{m_2}{\varepsilon_1})\sin\theta
-\frac{m_1\varepsilon_1\sin\theta}{s^2}\\
&+&(1+\frac{m_2}{\varepsilon_1^2})\varepsilon_2\sin\theta
-\frac{m_1}{s^3}(\varepsilon_2s\sin\theta-\varepsilon_1^2\sin2\theta),\nonumber
\end{eqnarray}
where $\varepsilon_1$ and $\varepsilon_2$ are given by
\begin{equation}
\varepsilon_1=\left(m_2\frac{m_1\cos2\theta\pm\sqrt{s^4-m_1^2\sin^22\theta}}
{s^2-m_1/s^2}\right)^{1/2},
\end{equation}
\begin{equation}
\varepsilon_2=-\frac{m_1\varepsilon_1^4(m_1\varepsilon_1^2\cos\theta+m_2s^2\cos3\theta)}
{m_2s^3(m_1\varepsilon_1^2\cos2\theta+m_2s^2)}.
\end{equation}
Here, $m_1$ and $m_2$ are the masses of lens star and planet,
respectively, and $s$ is the projected separation between
them normalized by the Einstein ring radius.

Hence, the planetary caustic size along the planet-star axis is
\begin{equation}\label{delx1}
\Delta
x=|x(0)-x(\pi)|=\frac{2\sqrt{m_2}}{s}\frac{m_1(2s^2+m_1+1)}{\sqrt{(s^4-m_1)(s^2+m_1)}}\;.
\end{equation}
If we choose $m_1=1$ and $m_2=q=m_2/m_1$ \citep[also see][]{han06},
then Equation \ref{delx1} becomes
\begin{equation}\label{delx}
\Delta x \simeq \frac{4\sqrt{q}}{s\sqrt{s^2-1}}\;.
\end{equation}
As shown in Equation \ref{delx}, the planetary caustic size $\Delta
x$ is proportional to $\sqrt{q}$ and inversely to $s^2$ when
$s\gg1$. Thus, a low mass planet with wide separation has a fairly
small perturbation region.

\section{FRACTIONAL DEVIATION MAP}

In Figure 1, we show the fractional deviation map for various ratios
of the source radius normalized by the planetary caustic size
$\rho_\star/\Delta x$, where  $\rho_\star=\theta_\star/\theta_{\rm
E}$ and $\theta_\star$ being the angular radius of a source star,
and normalized separation $s$ by the Einstein ring radius
$\theta_{\rm E}$. The ratio of the normalized source radius and
separation are denoted on the top of each column and on the
right-hand side of each row, respectively.

The fractional deviation $\sigma$ is defined  by
\begin{equation}
\sigma\equiv\frac{A-A_{\rm P}}{A_{\rm P}},
\end{equation}
where the magnification of the planetary lensing $A$ results from a
planet with its host star, and $A_{\rm P}$ from a planet alone, like
the case of a free-floating planet. Note that $A_{\rm P}$ is scaled
by the Einstein ring radius of the planet itself,
$\theta_{\rm{E,P}}$, which is related to the Einstein ring radius
for the total mass of the star-planet system, $\theta_{\rm E}$, as
$\theta_{\rm{E,P}}=\theta_{\rm E}\sqrt{q/(1+q)}$. The position of
the peak magnification of the planet itself, $\hat{s}$, is also
adjusted to match the center of a planetary caustic of the
star-planet system, i.e., $\hat{s}=s-1/s$.

Since a low mass planet with wide separation has a fairly small
perturbation region, the finite source effect is crucial. We
construct the magnification maps using the inverse ray-shooting
method to include the finite source effect
\citep{sch86,kay86,wam97}. In addition to the finite source effect,
we also take into account the limb darkening effect by the surface
of source star. On the surface of the source star, the specific
intensity $I$ with the flux $F$ and the limb darkening coefficient
of $\Gamma$ is given by
\begin{equation}
I=\frac{F}{\pi\theta_\star^2}[1-\Gamma(1-1.5\cos\phi)],
\end{equation}
where $\phi$ is the angle between the normal direction to the star's
surface and the direction toward the observer \citep{mil21,an02}. In
this particular study, we adopt a fixed value of  $\Gamma$ for all
source stars, i.e.,  $\Gamma=0.5$. We also assume that the lens
system with the mass ratio of $1\times10^{-3}$ is located at $D_l=6$
kpc and the source star is at $D_s=8$ kpc.

The panels in Figure 1 are divided by thick black lines with arrows
to outline in the parameter space the domains where regions with the
amplitude of fractional deviations as indicated by the arrows can be
found in the deviation maps. Note that the contours are drawn at the
levels of $\sigma=\pm1\%$, $\pm5\%$, $\pm10\%$, $\pm15\%$, and
$\pm20\%$ as shown in the scale-bar. As one may expect, the
fractional deviation decreases as the ratio of the source radius to
the planetary caustic size and/or the projected separation
increases.

\section{LENSING BY A WIDE-SEPARATION PLANET}

In Figure 2, we show the upper limits of normalized source radius
$\rho_\star/\Delta x$ that allow to detect the signature of the host
star as a function of the separation when a threshold
$\sigma_{\rm{th}}$ is given. Different symbols represent different
threshold  $\sigma_{\rm{th}}$, that is, $\sigma_{\rm{th}}=$ 5$\%$,
10$\%$, 15$\%$, and 20$\%$ for filled circles, triangles, squares,
and pentagons, respectively. The error bars mean the uncertainty on
the determination of the upper limit of $\rho_\star/\Delta x$. When
$s$ becomes much larger than unity, the magnification pattern can be
approximated by the \citet{cha79,cha84} lensing, and thus
$\rho_\star/\Delta x$ is nearly constant independent of $s$.
However, the planetary caustic becomes asymmetric and the size of
the planetary caustic also becomes bigger as $s$ approaches unity.
This is why $\rho_\star/\Delta x$ rapidly increases as $s$ gets
close to unity since the major perturbation regions around the cusps
of the planetary caustic are located outside of the Einstein ring
size of planet mass.

Another point to note is that this upper limit on the source-caustic
size ratio can be approximated by a fitting function,
\begin{equation}\label{fit}
\frac{\rho_\star}{\Delta
x}(s)=\frac{C_1}{s\sqrt{s^2-1}}+C_2\;\;\textrm{for}\;s\geq3,
\end{equation}
with $C_1\simeq0.3/\sigma_{\rm{th}}\sqrt{\sigma_{\rm{th}}}$ and
$C_2\simeq0.21/\sqrt{\sigma_{\rm{th}}}$, respectively, which is
represented by solid curves. What we have seen here has interesting
implications. For instance, this plot can be read to see how the
finite source effect constrain the range of the projected separation
when the detection threshold is given. In other words, one may
obtain for a given source radius and a planet-star mass ratio the
upper limit of the star-planet separation that allows the detection
of caustic structure due to a host star. Specifically speaking,
combining Equations \ref{delx} and \ref{fit}, one obtains
\begin{equation}\label{smax}
s\leq\left(\frac{1+\sqrt{1+4R^{-2}}}{2}\right)^{1/2},
\end{equation}
where
\begin{equation}
R\equiv\frac{-C_2+\sqrt{C_2^2+C_1\rho_\star / q^{1/2}}}{2C_1}.
\end{equation}

In Figure 3, we show the upper limit of the star-planet separation
to detect the boundness of the wide-separation planet as a function
of the source radius and the mass of the planet (left panel),
assuming for $\sigma_{\rm{th}}=5\%$, and of the detection threshold
of fractional deviation and the mass of the planet (right panel),
assuming for $R_\star=1\;R_{\odot}$. Dotted contours and gray scales
represent the upper limit of the separation in units of AU. The
planet mass $M_{\rm{p}}$ is given in a log scale in units of the
solar mass $M_{\odot}$, the source radius $R_\star$ in the left
panel is in units of the solar radius $R_{\odot}$. In this specific
example, we set $D_l=6$ kpc, $D_s=8$ kpc, and the lens mass of 0.5
$M_{\odot}$.

According to the left panel, the maximum separation one can tell
whether a planetary microlensing feature is caused by a bound planet
becomes larger as the radius of the source star becomes smaller
and/or the planet mass larger. In other words, as the source star is
large and the finite source effect becomes important, one can only
detect the existence of its host star when the planet is massive and
planet-star separation is small. When 5$\%$ of the detection
threshold is assumed, for a source star with the radius of $\sim1$
$R_{\odot}$, an Earth-mass planet ($\log M_{\rm{p}}=-5.5$
$M_{\odot}$) and a Jupiter-mass planet ($\log M_{\rm{p}}=-3.0$
$M_{\odot}$) can be recognized of its boundness when it is within
the separation range of $\sim10$ AU and $\sim30$ AU, respectively.
Similarly, according to the right panel, the upper limit of the
separation becomes larger as the detection threshold becomes lower.
The photometric precision of KMTNet project is expected to meet
$1\%$ at 21 magnitude in \textit{V} band on 10 minute monitoring
frequency and that of \textit{WFIRST} mission is $\leq1\%$ at 20.5
magnitude in \textit{J} band on $\leq15$ minute sampling cadence.
Therefore, we expect that the next-generation microlensing
experiments with high survey monitoring frequency and accurate
photometry will discover wide-separation planets with various
separations as analyzed in this study.

In Figure 4, we compare the separation ranges to detect the
boundness of the wide-separation planet through the channel of the
central and planetary caustics. The black curves represent the upper
limits of the separation range induced by the planetary caustic for
various source radii $R_\star=$ $1\;R_{\odot}$, $3\;R_{\odot}$,
$5\;R_{\odot}$, and $7\;R_{\odot}$, while the gray curves represent
by the central caustic that we calculate following \citet{han09a}.
Results are given both in units of AU and $\theta_{\rm{E}}$. We take
the same values of parameters as in Figure 3 for the distances of
lens and source, and the lens mass, and assuming 5$\%$ of the
detection threshold. The separation range to detect the boundness of
the wide-separation planet induced by the planetary caustic is wider
than that induced by the central caustic. What it means is that,
when the microlensing light curve induced by the planetary caustic
happens to be analyzed, one may afford to detect the boundness of
the wide-separation planet farther than when that caused by the
central caustic is analyzed.

\section{CONCLUSION}

The unbiased spatial distribution of extrasolar planets with respect
to host stars is crucial in studying the planet formation processes.
Massive and/or close extrasolar planets are likely to be detected by
a commonly employed method. On the other hand, the microlensing
technique can be sensitive to Earth-like and/or wide-separation
planets. To obtain the spatial distribution of extrasolar planets
without contaminating by free-floating planets, it is important to
characterize the planetary microlensing light curves whether they
are caused by a wide-separation planet or a free-floating planet.

Here, we analyze the condition in terms of lensing parameters,
including the size of source stars, under which signatures of the
host star in the planetary microlensing light curve can be detected.
By constructing the fractional deviation maps at various positions
in the space of microlensing parameters, we have obtained the upper
limits of  $\rho_\star/\Delta x$ that allow to detect the signature
of the host star as a function of the separation when a threshold is
given. We confirm that when $s\gg1$ the \citet{cha79,cha84} lensing
well-approximate what we have found. We also note that a simple
analytical function can be fit, as given in Equation \ref{fit}. This
relation further leads one to a simple analytic condition for the
star-planet separation to detect the boundness of wide-separation
planets as a function of the mass ratio and source radius, as shown
in Figure 3. Finally, we have compared the separation ranges to
detect the boundness of the wide-separation planet through the
channel of the central and planetary caustics. As a result, we
conclude that when the microlensing light curve caused by the
planetary caustic happens to be analyzed, one may afford to support
the boundness of the wide-separation planet farther than when that
caused by the central caustic is analyzed. Therefore, we conclude
that the next-generation microlensing experiments with high survey
monitoring frequency are expected to add the number of
wide-separation planets through the channel of the planetary
caustic.

\section*{Acknowledgments}
We thank the anonymous referee for critical comments which clarify
and improve the original version of the manuscript. This research
was supported by Basic Science Research Program through the National
Research Foundation of Korea (NRF) funded by the Ministry of
Education, Science and Technology (2012R1A6A3A01013815) for YHR and
(2012R1A1A4A01013596) for MGP. MGP was also supported by the
National Research Foundation of Korea to the Center for Galaxy
Evolution Research (NO.2010-0027910). HYC was supported by the
National Research Foundation of Korea Grant funded by the Korean
government (NRF-2011-0008123).

\clearpage

\begin{figure*}
\begin{minipage}[hpt]{17cm}
\begin{center}
\includegraphics[scale=1.0,angle=0,clip=true]{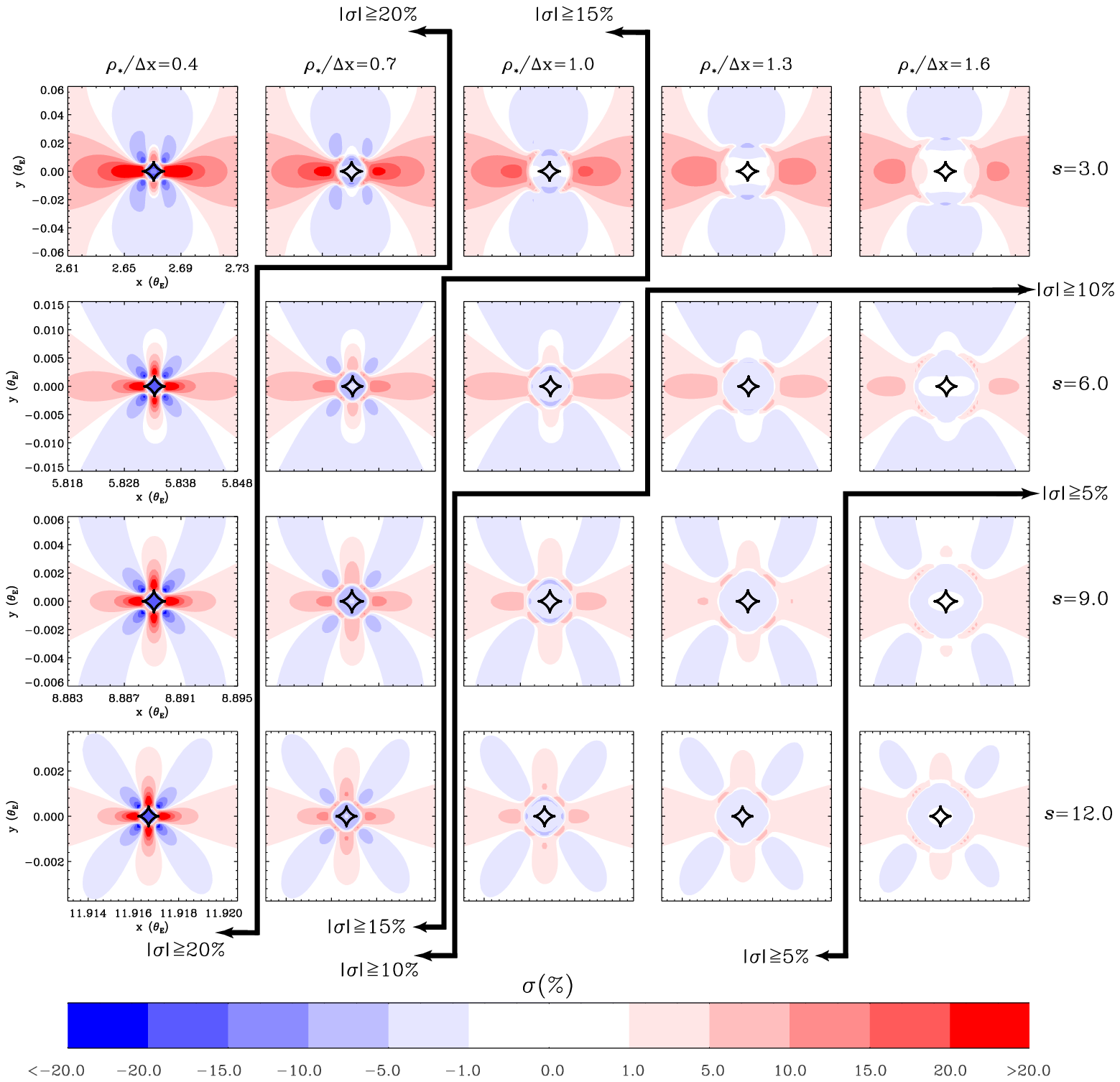}
\caption{Maps of the fractional deviation in  magnifications for
various ratios of the source radius normalized by the planetary
caustic size $\rho_\star/\Delta x$, where
$\rho_\star=\theta_\star/\theta_{\rm E}$ and $\theta_\star$ being
the angular radius of a source star, and normalized separation $s$
by the Einstein ring radius $\theta_{\rm E}$. The ratio of the
source radius to the planetary caustic size $\rho_\star/\Delta x$
and the separation $s$ are marked on the top of each column and the
right-hand side of each row, respectively. The panels are divided by
thick black lines with arrows to outline in the parameter space the
domains where regions with the amplitude of fractional deviations as
indicated by the arrows can be found in the deviation maps. In each
map, the contours are the levels of $\sigma=\pm1\%$, $\pm5\%$,
$\pm10\%$, $\pm15\%$, and $\pm20\%$, and the colors of blue and red
tones are the regions with negative and positive $\sigma$,
respectively, where lighter shades represent smaller $|\sigma|$
levels. We show $\sigma$ levels corresponding to each color in the
scale-bar. The black curve at the center in each panel represents
the caustic shape. In this particular plot, we set the Galactic
distances of lens $D_l=6$ kpc and that of source star $D_s=8$ kpc.
The lens star is located at the origin of coordination, i.e.,
($x,y$)=(0,0).}
\end{center}
\end{minipage}
\end{figure*}

\begin{figure*}
\begin{minipage}[hpt]{17cm}
\begin{center}
\includegraphics[scale=1.0,angle=0,clip=true]{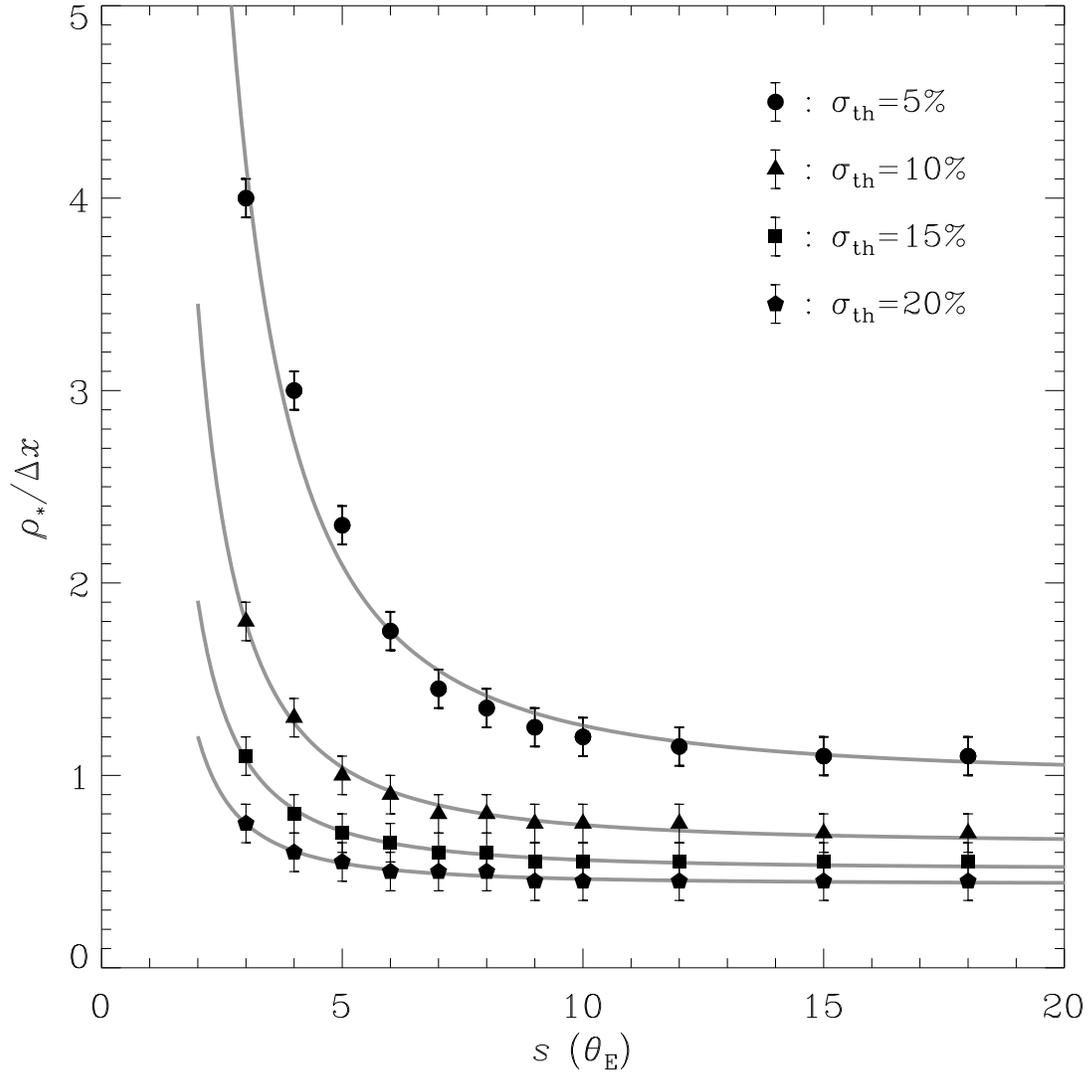}
\caption{Upper limits on the ratio of the source radius to the
planetary caustic size for $\sigma_{\rm{th}}=$ 5$\%$, 10$\%$,
15$\%$, and 20$\%$. The filled circles, triangles, squares, and
pentagons represent the maximum values of $\rho_\star/\Delta{x}$ as
a function of $s$ for different thresholds  $\sigma_{\rm{th}}$, that
is, $\sigma_{\rm{th}}=$ 5$\%$, 10$\%$, 15$\%$, and 20$\%$,
respectively. The error bars mean the uncertainty on the
determination of the upper limit of $\rho_\star/\Delta x$. The black
curve is the fitting function, that is, Equation \ref{fit}. See text
for detailed discussions of this plot.}
\end{center}
\end{minipage}
\end{figure*}

\begin{figure*}
\begin{minipage}[hpt]{17cm}
\begin{center}
{\includegraphics[scale=0.5,angle=0,clip=true]{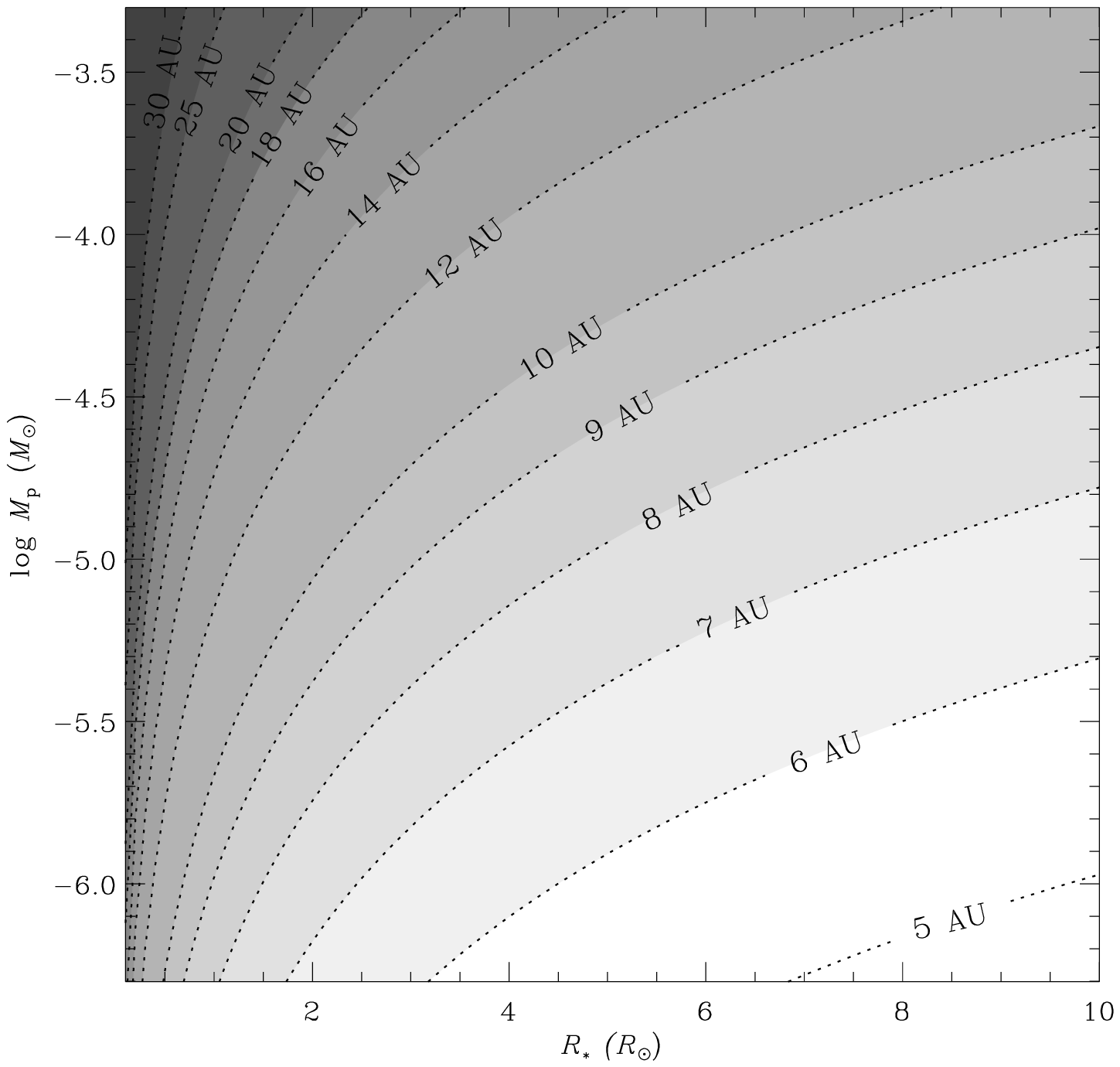}}
{\includegraphics[scale=0.5,angle=0,clip=true]{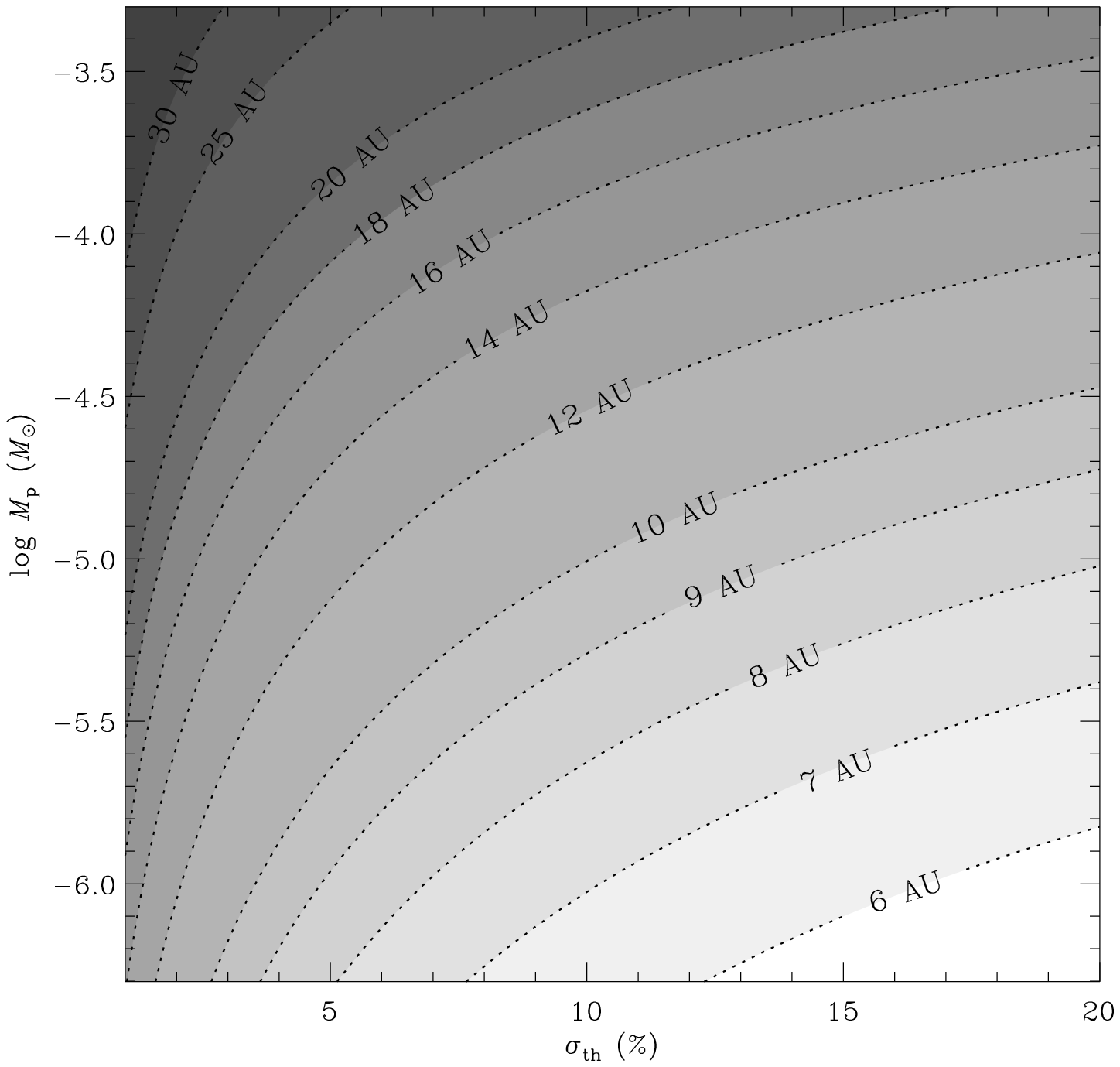}}
\caption{Upper limit of the star-planet separation to detect the
boundness of the wide-separation planet as a function of the source
radius and the mass of the planet (\textit{left panel}), assuming
for $\sigma_{\rm{th}}=5\%$, and as a function of the detection
threshold of fractional deviation and the mass of the planet
(\textit{right panel}), assuming for $R_\star=1\;R_{\odot}$. Dotted
contours and gray scales represent the upper limit of the separation
in units of AU. The employed parameters in this example are $D_l=6$
kpc, $D_s=8$ kpc, and lens mass of 0.5 $M_{\odot}$.}
\end{center}
\end{minipage}
\end{figure*}

\begin{figure*}
\begin{minipage}[hpt]{17cm}
\begin{center}
\includegraphics[scale=1.0,angle=0,clip=true]{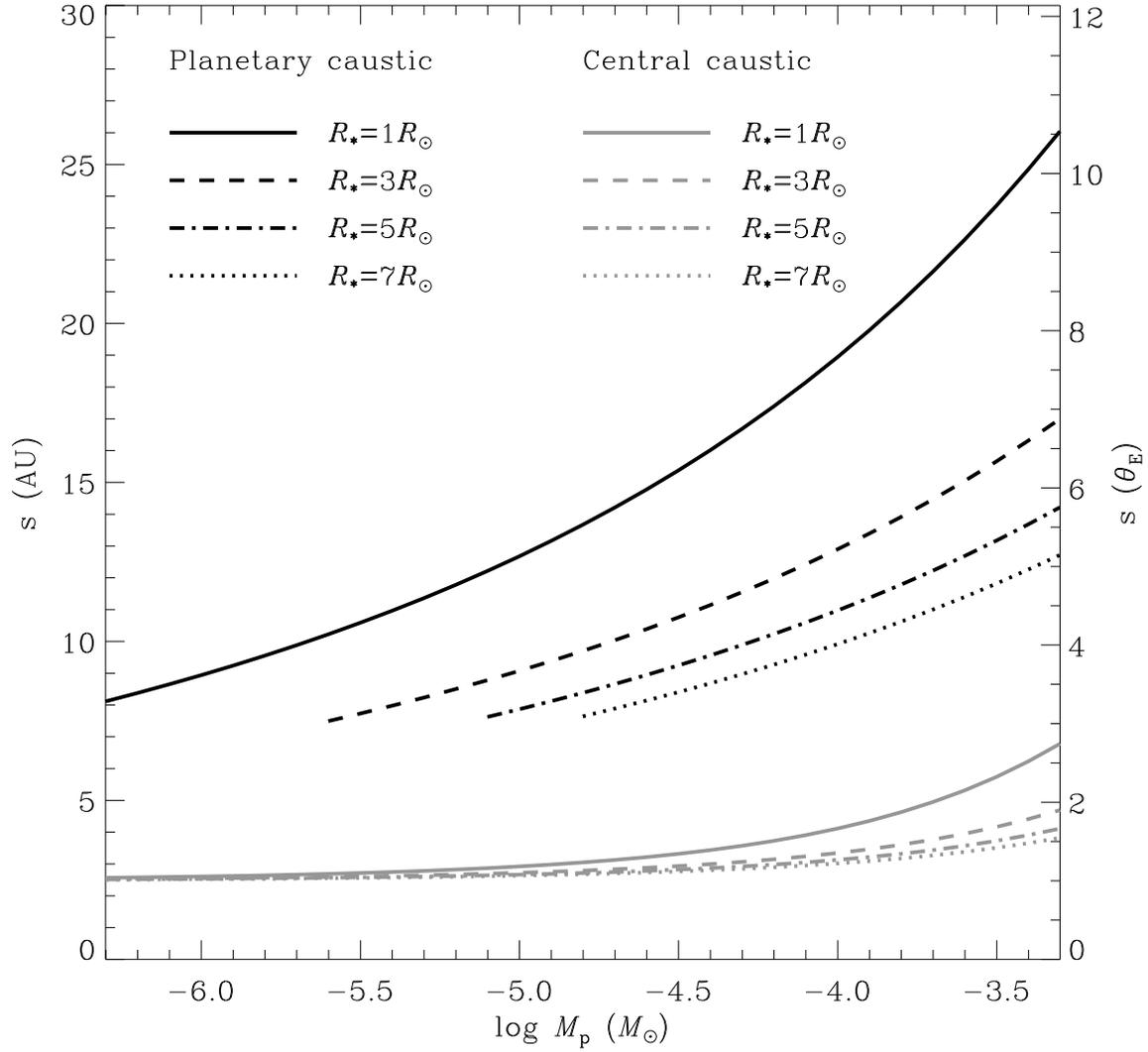}
\caption{Comparison of the separation ranges to detect the boundness
of the wide-separation planet through the channel of the central
caustics with the planetary caustics. The upper limits of the
separation ranges are plotted with the black and gray curves for the
planetary and central caustics, respectively. We assume that the the
distances of lens and source and the lens mass are $D_l=6$ kpc,
$D_s=8$ kpc, and 0.5 $M_{\odot}$, and the detection threshold is
5$\%$.}
\end{center}
\end{minipage}
\end{figure*}

\label{lastpage}

\end{document}